\documentclass{article}

\usepackage{PRIMEarxiv}

\usepackage[utf8]{inputenc} 
\usepackage[T1]{fontenc}    
\usepackage{hyperref}       
\usepackage{url}            
\usepackage{booktabs}       
\usepackage{amsfonts}       
\usepackage{nicefrac}       
\usepackage{microtype}
\usepackage{amsmath}
\usepackage{lipsum}
\usepackage{fancyhdr}       
\usepackage{graphicx}       
\graphicspath{{media/}}     

\title{The Skeletal Trap: Mapping Spatial Inequality and Ghost Stops in Ankara’s Transit Network

}

\author{
Elifnaz Kancan \\
Independent Researcher \\
Ankara, Turkey \\
\texttt{en.kancan@gmail.com}
}

\begin{document}
\maketitle

\begin{abstract}
Ankara’s public transport crisis is commonly framed as a shortage of buses or operational inefficiency. This study argues that the problem is fundamentally morphological and structural. The city’s leapfrog urban expansion has produced fragmented peripheral clusters disconnected from a rigid, center-oriented bus network. As a result, demand remains intensely concentrated along the Kızılay–Ulus axis and western corridors, while peripheral districts experience either chronic under-service or enforced transfer dependency. The deficiency is therefore not merely quantitative but rooted in the misalignment between urban macroform and network architecture.

The empirical analysis draws on a 173-day operational dataset derived from route-level passenger and trip reports published by EGO under the former “Transparent Ankara” initiative. To overcome the absence of stop-level geospatial data, a Connectivity-Based Weighted Distribution Model reallocates passenger volumes to 1 km × 1 km grid cells using network centrality. The findings reveal persistent center–periphery asymmetries, structural bottlenecks, and spatially embedded accessibility inequalities.
\end{abstract}

\keywords{Ankara \and Grid-based Spatial Analysis \and Transportation \and Service-Demand Mismatch}

\section{Introduction}
Recent complaints on social media and in local news about Ankara’s public transport show a system under strain. Overcrowded vehicles, long waiting times in outer districts, and tired daily passengers point to problems in operations or a lack of vehicles \cite{yilmaz2023ankara,yilmaz2023ego,sabah2026ankara,evrensel2026ankara}. While the public demands \textit{"more buses"}, this research argues that these daily complaints are signs of a long-term planning crisis in transportation and urban planning.

\hspace{1cm}This structural mismatch has historical roots. Rapid migration beginning in the 1950s quickly overwhelmed Ankara’s existing transport system. By the late 1960s, the public sector's share in transport had dropped significantly, forcing a reliance not only on informal mobility solutions such as \textit{dolmuş} but also on institution-based personnel shuttles to bridge the gap left by insufficient municipal services \cite{Oncu2009}.\footnote{A dolmuş is a shared taxi or minibus in Turkey that follows a semi-fixed route and departs when full, providing flexible, informal public transport.}  Early transportation studies in the 1970s proposed rail-based alternatives, yet pioneering projects like the 1972 Sofretu plan were rejected by central authorities due to financial constraints and political concerns over foreign technological dependence \cite{cubuk2003ankara, Oncu2009}. While the 1980s saw the institutionalization of private transit operators to compensate for the lack of investment, the most drastic morphological shift occurred post-1994. The municipal administration explicitly prioritized private vehicle mobility, diverting resources from maintaining the bus fleet—which actually shrank during this period—toward the construction of multi-level interchanges  and signal-free road corridors \cite{Oncu2009}. This policy fragmented the pedestrian fabric and stalled critical rail investments, turning metro projects into prolonged \textit{"construction sagas"}  that failed to keep pace with the city's leapfrog growth \cite{batuman2013city, Gunay2012}. As a result, Ankara’s present transit system reflects decades of planning that prioritized suburban growth and automobile circulation over an integrated, adaptive network \cite{Tekeli2012}.

To understand why the bus network fails to cover the city effectively, one must look at how Ankara has grown. In his analysis of the city’s form,  Günay argues that Ankara did not expand as a  continuous \textit{"oil stain" }radiating from the center to the periphery, as observed in traditional urban ecological models \cite{Gunay2012,LightPollutionMap_Ankara}. Instead, driven by speculative real estate markets and fragmented planning decisions, the city exhibited a \textit{"leapfrog"} development pattern \cite{Gunay2012,Tekeli2012}. This disjointed morphology created isolated clusters of residential and industrial areas on the fringe, separated by vast and undeveloped voids.

\begin{figure}
    \centering
    \includegraphics[width=1\linewidth]{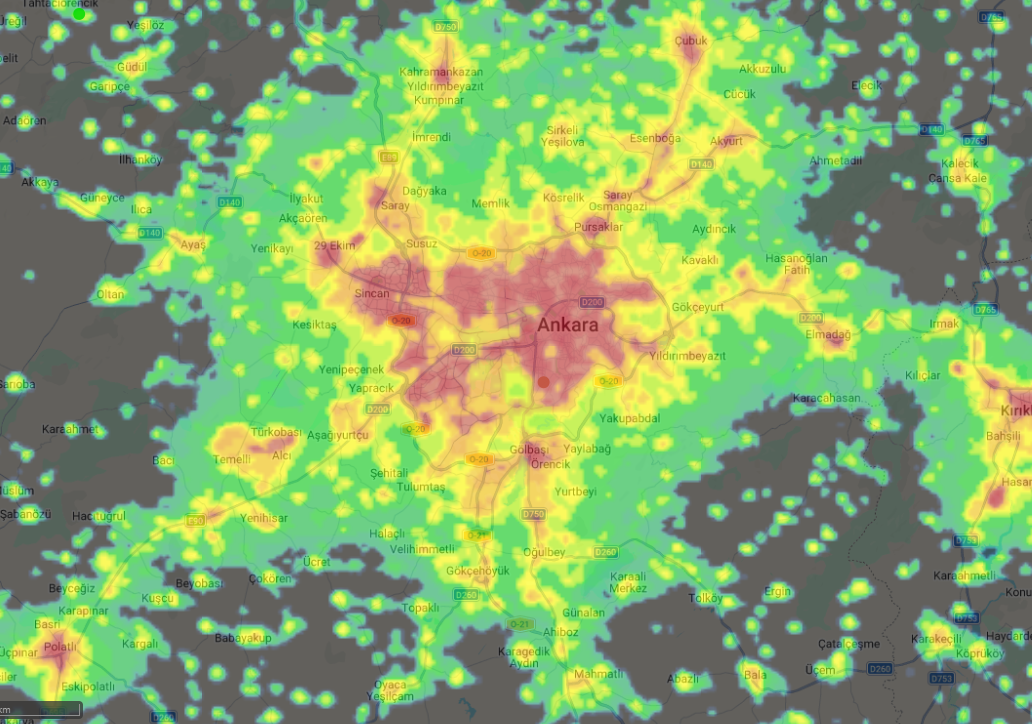}
    \caption{\textbf{Urban Activity Proxy (Nighttime Lights):} Heatmap illustrating Ankara's leapfrog expansion and peripheral clusters outpacing traditional transit infrastructure  in 2025.\cite{LightPollutionMap_Ankara}.}
    \label{fig:lightpol}
\end{figure}

\hspace{1cm}The structural deficiency of the bus service is further exacerbated by the organization of the network itself. Ankara’s transport policy has historically prioritized private vehicle ownership and highway speed over integrated public transit \cite{Erdal2012}. The existing bus network operated by EGO (Ankara Electricity, Gas and Bus Operations Organization) is not organized as an integrated mesh. It consists of separate lines that often remain within administrative district boundaries. This outcome reflects the long-term planning paradigm described by Batuman, in which transportation investments consistently favored vehicular mobility and suburban accessibility over integrated public transport systems \cite{batuman2013city}. The planning of the transporation forces a systemic \textit{"transfer necessity"} , compelling populations living in the periphery to travel to the city center (Kızılay-Ulus axis) even for orbital trips between districts \cite{Erdal2012}. This radial dependence creates a bottleneck at the center and leaves the peripheral \textit{"leapfrog"} zones underserved, turning the act of transportation into a mechanism of spatial segregation.

\hspace{1cm}This misalignment generates accessibility inequalities. Ilhan Tekeli defines range not just as physical distance, but as \textit{the capacity of a citizen to utilize the city's opportunities}; a capacity that is severely restricted for those dependent on public transit \cite{Tekeli2012}.Some of the  distinct scenarios reveal that low-income workers living in peripheral areas like Sincan or Karapürçek spend approximately 4.5 hours daily in transit due to inefficient routes and mandatory transfers \cite{Erdal2012}. This \textit{"time poverty"} effectively detaches these citizens from the \textit{"social space"} of the city, reducing them to mere labor power in circulation and transforming mobility from a right into a privilege \cite{Erdal2012,Balkanay2012}. Thus, the bus service deficiency in Ankara is fundamentally a manifestation of spatial injustice, produced by the friction between a sprawling urban form and an obsolete network design.

\section{Methodology and Analytical Framework}
This study presents quantitative analysis aimed at detecting supply-demand imbalances and operational instabilities  within Ankara's public transport network. The research relies on a dataset covering a 173-day monitoring period (between December 25, 2023, and October 20, 2024), derived from daily passenger and trip reports published by the EGO General Directorate. Although these records were harvested via web scraping techniques, they were originally disseminated under the \textit{"Transparent Ankara"} (Şeffaf Ankara) open data initiative. Data was obtained from the EGO General Directorate's official records. After October 20, 2024, bus data were discontinued and metro data were shared only intermittently.\footnote{\url{https://www.ego.gov.tr/tr/sayfa/2297/gunluk-yolcu-ve-sefer-sayilari}, accessed December 2025.}
However, this initiative presented significant utility challenges: the data lacked essential geolocation components (GTFS) and hourly temporal resolution, providing only daily route-level aggregates. Furthermore, as this open data source has subsequently been removed from public access, the harvested dataset represents a unique, preserved cross-section of the network's operational history. To reconstruct the missing spatial dimension for 50,424 unique stops and enable spatial analysis, a hierarchical geocoding pipeline was developed, prioritizing high-precision commercial APIs while utilizing algorithmic estimation to fill critical gaps in the network periphery.

The coordinate retrieval process was structured across three primary tiers:

\begin{itemize}
  \item \textbf{Google Maps API (73.6\% -- 37,122 stops):} The Google Maps Platform served as the main reference source. The algorithm queried stop names using three address formats such as ``Ankara, \{address\}'' and ``\{stop\_name\}, Ankara'' to improve matching rates and obtain high spatial accuracy for the main network.

  \item \textbf{Algorithmic Interpolation (14.3\% -- 7,190 stops):} For stops not found in commercial or open databases, a mathematical interpolation method estimated coordinates by calculating a weighted midpoint between known previous and next stops on a route. This step was important for mapping newly developing peripheral areas where digital map coverage is limited.

  \item \textbf{OpenStreetMap Integration (2.2\% -- 1,117 stops):} OpenStreetMap data were used as a secondary validation layer. Although 2,693 stops were tagged as \texttt{highway=bus\_stop} in the Ankara region, the dataset was used only to fill specific gaps when the main API failed, providing final coordinates for 1,117 stops.
\end{itemize}

\hspace{1cm}To overcome the absence of coordinate data and stop-level granularity, the study implements a mathematical model that divides the city into standardized \textit{ 1 km × 1 km} grid cells. It resolves the spatial ambiguity by distributing passenger loads based on a \textit{Connectivity-Based Weighted Distribution model}, which assigns higher probabilities to network hubs rather than using uniform distribution. To address the limitations imposed by the absence of stop-level boarding data (data scarcity), this study rejects the Naive Uniform Distribution model, which erroneously assigns equal passenger probability to all stops regardless of their urban context. Instead, a Connectivity-Based Weighted Distribution Model is developed. This approach operationalizes Erdal’s (2012) concept of \textit{"transfer necessity",} which posits that Ankara’s transport morphology is strictly radial and center-dependent, forcing passenger accumulation at specific transfer hubs rather than dispersing it evenly across the network.

\hspace{1cm}The model distributes the known total route passenger count, $P_{\text{total}}$, to individual stops based on their network centrality, calculated through the following four-step formulation .

\paragraph{Step 1: Stop Connectivity Score ($C_s$).}
First, the network centrality of each stop is quantified. The connectivity score is defined as the cardinality of the set of unique routes passing through stop $s$:

\begin{equation}
C_s = \left| \left\{ r \in R : s \in S_r \right\} \right|
\end{equation}

where $R$ is the set of all routes and $S_r$ is the set of stops on route $r$.
Operational data reveals a stark hierarchy in Ankara's network: central ``Hub'' stops (e.g., K{\i}z{\i}lay/Ulus axis) exhibit connectivity scores as high as $C_s = 84$, while peripheral stops in  fragmented development zones often possess a score of $C_s = 1$, indicating extreme isolation .

\paragraph{Step 2: Route Aggregate Connectivity ($C_{\text{total}}(r)$).}
The total ``access power'' of a specific route $r$ is calculated by summing the connectivity scores of all stops along its trajectory:
\begin{equation}
C_{\text{total}}(r) = \sum_{s \in S_r} C_s
\end{equation}

\paragraph{Step 3: Stop Weight Coefficient ($W_s(r)$).}
The probability of passenger activity at stop $s$ is derived as a ratio of its individual connectivity to the route's total connectivity. This coefficient represents the stop's relative ``pull'' force within the network topology:
\begin{equation}
W_s(r) = \frac{C_s}{C_{\text{total}}(r)}
\end{equation}

Empirical results show that a central hub attracts approximately $60\%$ of a route's total load ($W_s \approx 0.60$),
whereas isolated peripheral stops receive only $0.2\%$ ($W_s \approx 0.002$), reflecting structural accumulation.

\paragraph{Step 4: Final Passenger Distribution ($P_s(r)$).}
Finally, the route-level passenger volume is distributed based on these calculated weights:
\begin{equation}
P_s(r) = P_{\text{total}}(r) \times W_s(r)
\end{equation}

\hspace{1cm}This weighted approach reduces the false positive rate in anomaly detection by $26$--$51\%$ compared to uniform distribution models, providing a statistically robust representation of Ankara.

\begin{figure}
    \centering
    \includegraphics[width=0.5\linewidth]{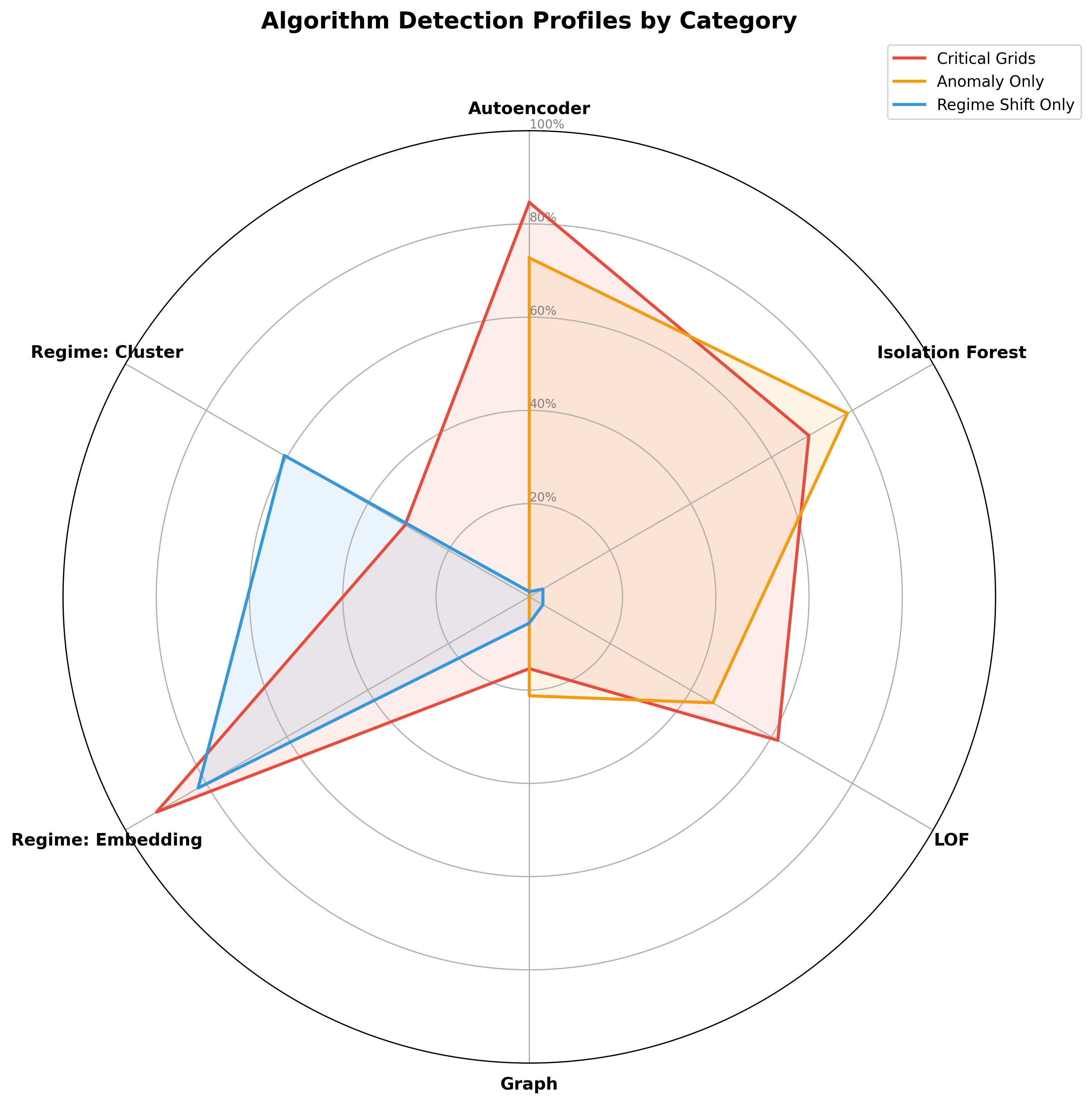}
    \caption{\textbf{Radar chart} illustrating the distinct algorithmic fingerprints of the three grid categories, demonstrating that Critical Grids (red) uniquely exhibit simultaneous failure across both static anomaly detection (left) and dynamic regime stability (right) axes, while Anomaly Only and Regime Shift Only categories display mutually exclusive failure patterns}
    \label{fig:radar}
\end{figure}

\hspace{1cm}The methodology classifies grid cells based on their behavioral fingerprints across six unsupervised dimensions, as shown in Figure \ref{fig:radar}. This multi-criteria structure prevents failures from being interpreted as artifacts of a single metric and instead captures distinct structural conditions. The first layer evaluates whether a grid deviates from the system’s static equilibrium using four anomaly detection algorithms: Autoencoder (AE), Isolation Forest (IF), Local Outlier Factor (LOF), and Graph-based Anomaly Detection (GAD). In the orange profile, grids labeled Anomaly Only expand strongly along these anomaly axes but remain close to the center on the Regime Shift dimensions. This indicates that their condition is persistent rather than episodic. Their failure is structural and stable, corresponding to chronic neglect, meaning areas that remain consistently underserved while maintaining internal stability.

\hspace{1cm}The second layer evaluates temporal stability, meaning the system’s ability to maintain a consistent operational identity, using Embedding Stability (ES) and Cluster Switching (CS) metrics. As seen in the blue profile, grids classified as Regime Shift Only expand on these axes while remaining low on the Anomaly axes. This shows areas that appear normal on average but experience high day-to-day variability, representing emerging sprawl, which are transitional zones where demand patterns have not yet stabilized. The most severe class, Critical Grids shown in the red profile, is defined by the intersection of these two layers. Covering both Anomaly and Regime axes, these grids represent structural collapse, areas where centralized transit planning fails to cope with spatiotemporal complexity and service reliability breaks down.

\subsection{Spatial Analysis \& Mapping}

\begin{figure}
    \centering
    \includegraphics[width=1\linewidth]{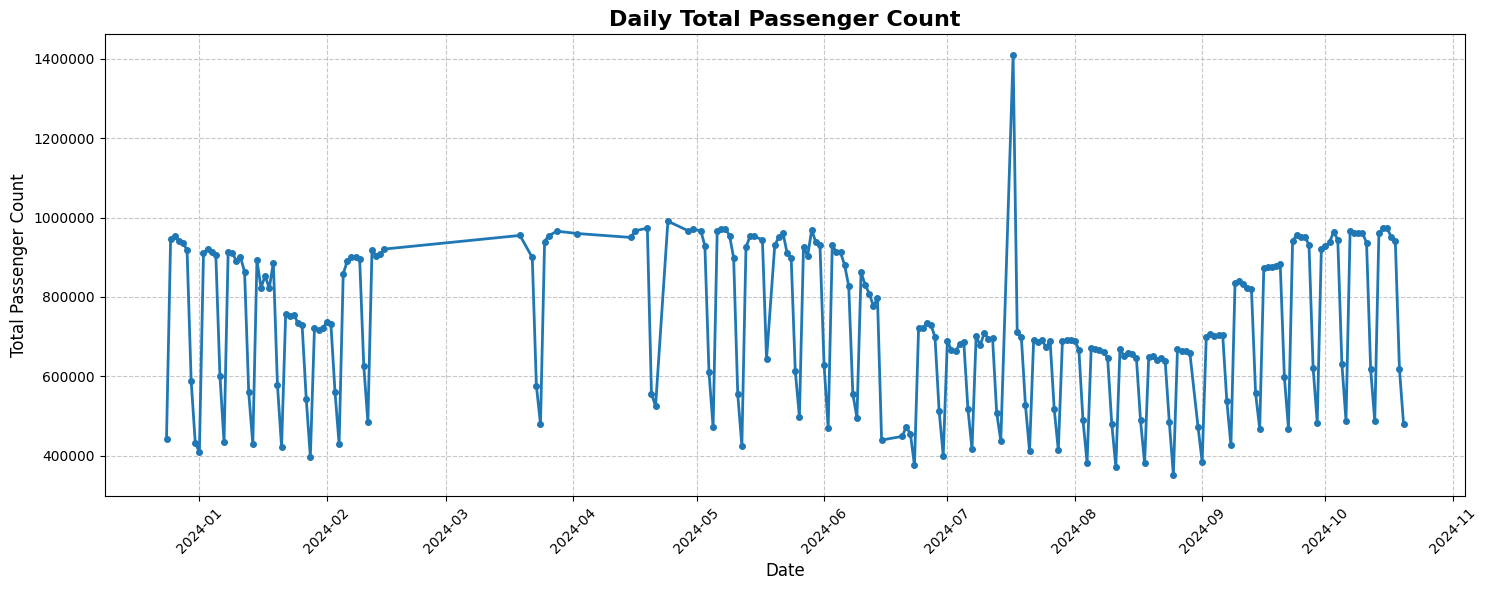}
    \caption{\textbf{Daily Passenger Count Trends and Temporal Anomalies}}
    \label{fig:placeholder}
\end{figure}

\begin{figure}
    \centering
    \includegraphics[width=1\linewidth]{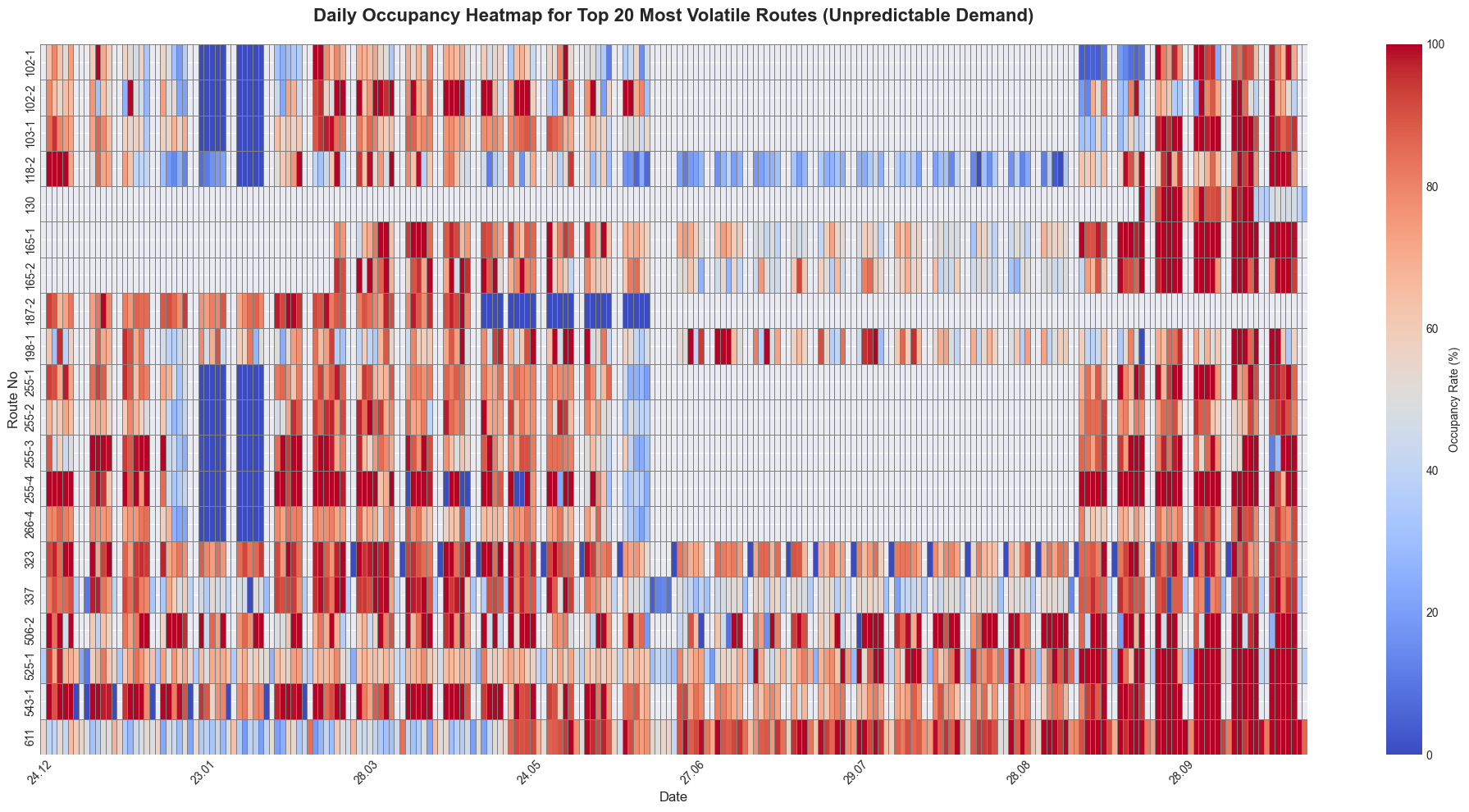}
    \caption{\textbf{Heatmap of daily occupancy rates for the 20 most volatile routes}, highlighting extreme fluctuations}
    \label{fig:volatile}
\end{figure}

The empirical foundation of this study is based on 112,930 route-level records covering a 173-day observation period. The analysis scripts and processed data are available in the project repository\footnote{\url{https://github.com/enkancan/ego-git}}.
Within this timeframe, the system operated across 691 unique stops, generating 2,069,345 daily trips and carrying a total of 170,722,567 passengers against a total available capacity of 327,776,765 seats. The system-wide occupancy rate stands at 52.1 percent, indicating that in aggregate terms the network appears neither critically overloaded nor severely underutilized. However, this aggregation conceals substantial internal problems. High-occupancy routes exceeding 85 percent account for only 4.7 percent of all records, whereas 67.1 percent fall within the medium occupancy band and 28.2 percent operate below 30 percent occupancy. The Figure \ref{fig:placeholder} reveals a rigid \textit{"sawtooth"} rhythm tied to institutional flows, where Sundays mark the absolute lowest demand points. This regular pattern collapses during late March and early April, with flat lines indicating data recording failures. A visible decrease appears in July, except for the apparent July 17 peak. This spike is not a real demand increase but a reporting artifact: data for July 15 and July 16 were not entered because of the holiday, and the passenger totals for three days were recorded on July 17, while the missing days appear as zero in the dataset. The route-level breakdown in Figure \ref{fig:volatile} further illuminates this instability, highlighting contrast between chronically saturated suburban lines—such as 543-1 and 611, and the service gaps and total data voids that characterize the network's most volatile corridors. This heatmap captures following school closures, alongside the structural volatility visible where specific routes are abruptly introduced or removed from service. Furthermore, the synchronized vertical blue stripes appearing across all routes suggest system-wide reporting anomalies where daily occupancy was likely defaulted to zero, mirroring the data recording failures identified in the aggregate temporal analysis.

\begin{figure}
    \centering
    \includegraphics[width=1\linewidth]{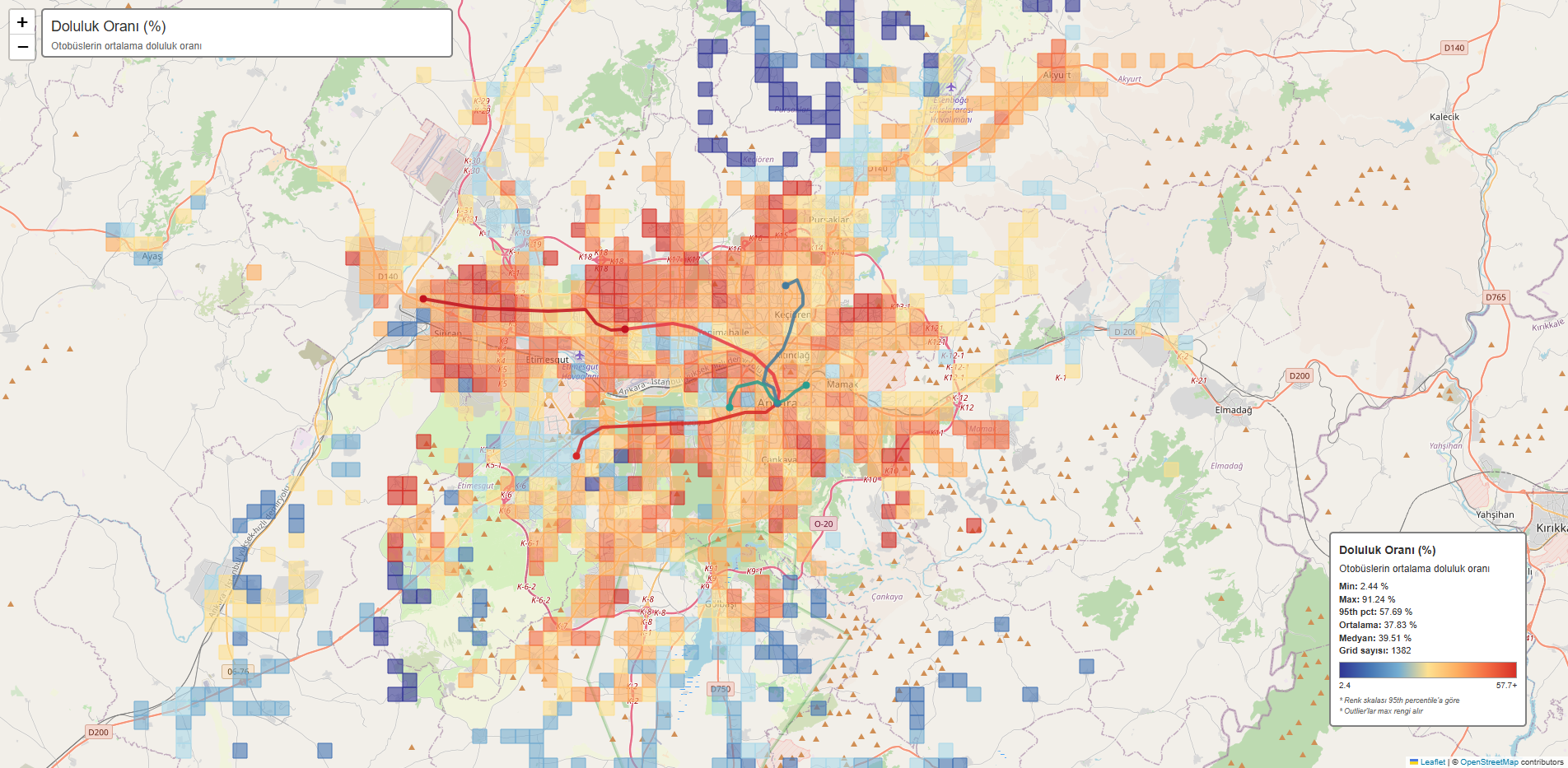}
    \caption{\textbf{Spatial Distribution of Average Occupancy Rates:} A grid-based visualization identifying extreme localized saturation (red zones) in peripheral districts such as Sincan and Batıkent, highlighting where demand exceeds available supply.}
    \label{fig:dolulukorani}
\end{figure}

\begin{figure}
    \centering
    \includegraphics[width=1\linewidth]{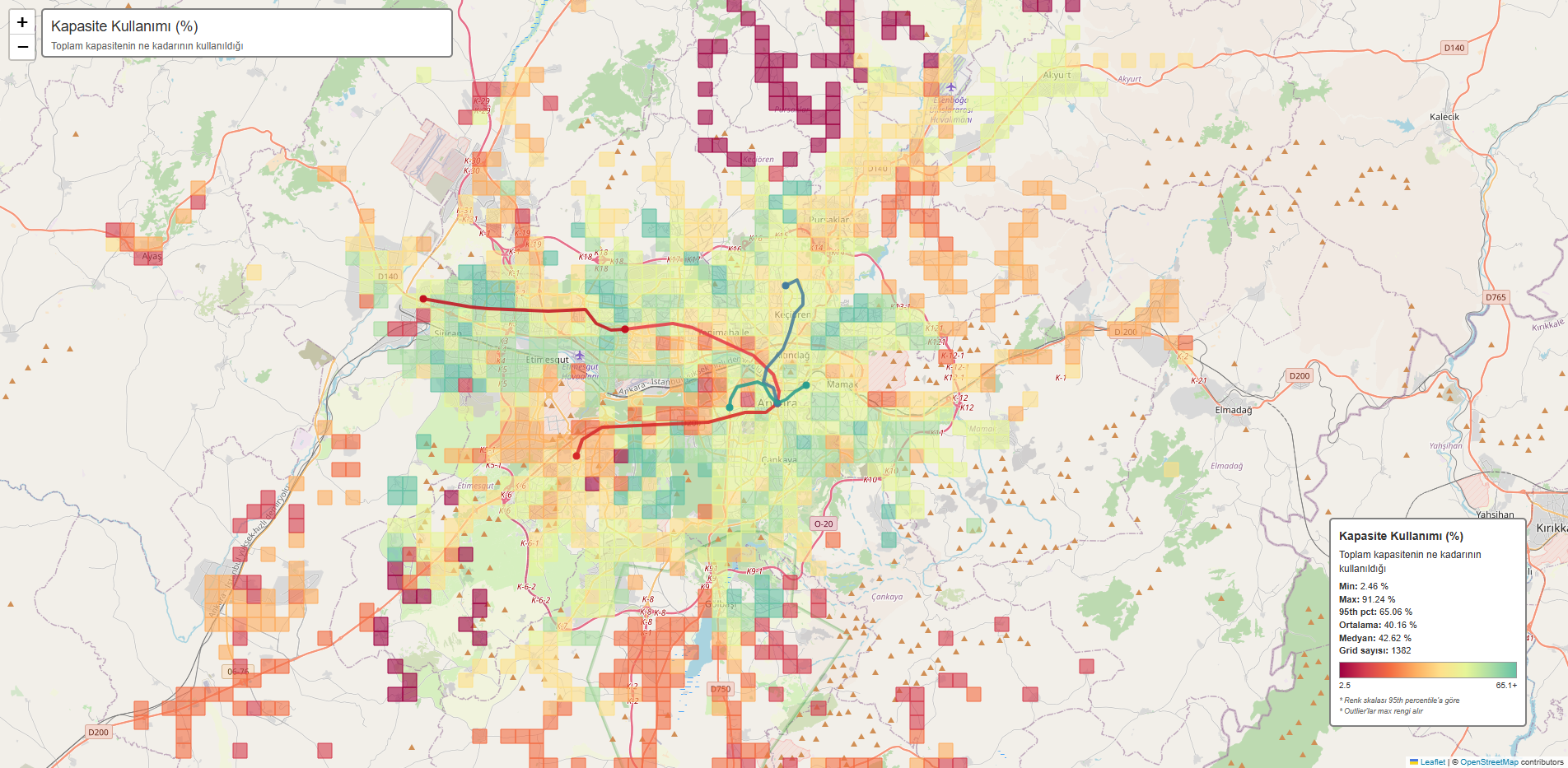}
    \caption{\textbf{Grid-based Capacity Utilization Analysis:} Mapping the percentage of used seat capacity across the network .}
    \label{fig:gridkapasite}
\end{figure}

\hspace{1cm}The morphological and structural misalignment of Ankara’s public transport network can be conceptualized as a \textit{"Skeletal Trap"}, a condition in which a rigid, center-oriented transit architecture fails to adapt to the city’s urban expansion in Figure \ref{fig:dolulukorani}. Conventional explanations emphasize fleet shortages, yet grid-based analysis of capacity and trip distribution reveals a different pattern: service frequency is disproportionately concentrated along the Kızılay–Ulus axis and major western corridors in Figure \ref{fig:gridkapasite}. This spatial concentration produces a contradictory pattern of forced peripheral demand. While central zones operate under sustained but manageable pressure, fragmented outer districts experience chronic under-service, where infrequent vehicles run at near-saturation levels. 

\begin{figure}
    \centering
    \includegraphics[width=1\linewidth]{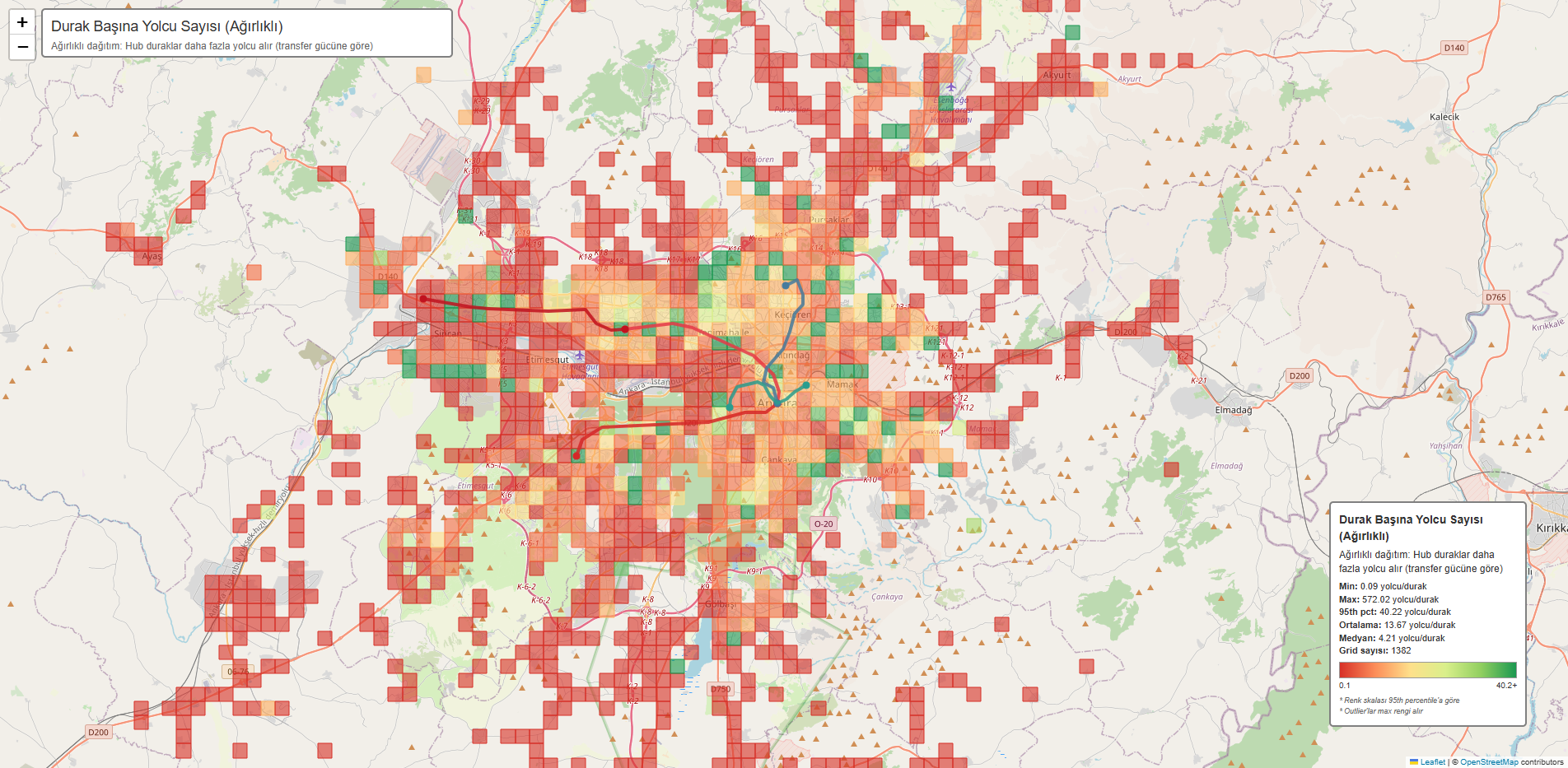}
    \caption{\textbf{Spatial density of passenger demand across Ankara’s bus network}, visualized as the average daily ridership per stop within 1 km² grid cells.}
    \label{fig:yolcustop}
\end{figure}

\hspace{1cm}The integration of the rail system further clarifies the rigidity of this structural condition. Metro lines function as a  backbone that organizes and constrains bus flows, reinforcing radial dependency rather than enabling distributed connectivity. This configuration institutionalizes transfer dependency by channeling peripheral passengers toward rail hubs regardless of their ultimate destinations, thereby intensifying pressure at specific interchange nodes in Figure \ref{fig:yolcustop}. The most acute manifestation of this dynamic appears at the terminal stations of the M1, M2, and M3 lines, including Koru and Törekent. At these endpoints, high-capacity rail infrastructure abruptly transfers demand to comparatively \textit{low-capacity feeder buses}, producing pronounced service mismatches and recurring bottlenecks. Instead of mitigating congestion, the rail network in these contexts amplifies spatial inequality by exposing the discontinuity between the rigid core and the fragmented periphery.

\begin{figure}
    \centering
    \includegraphics[width=1\linewidth]{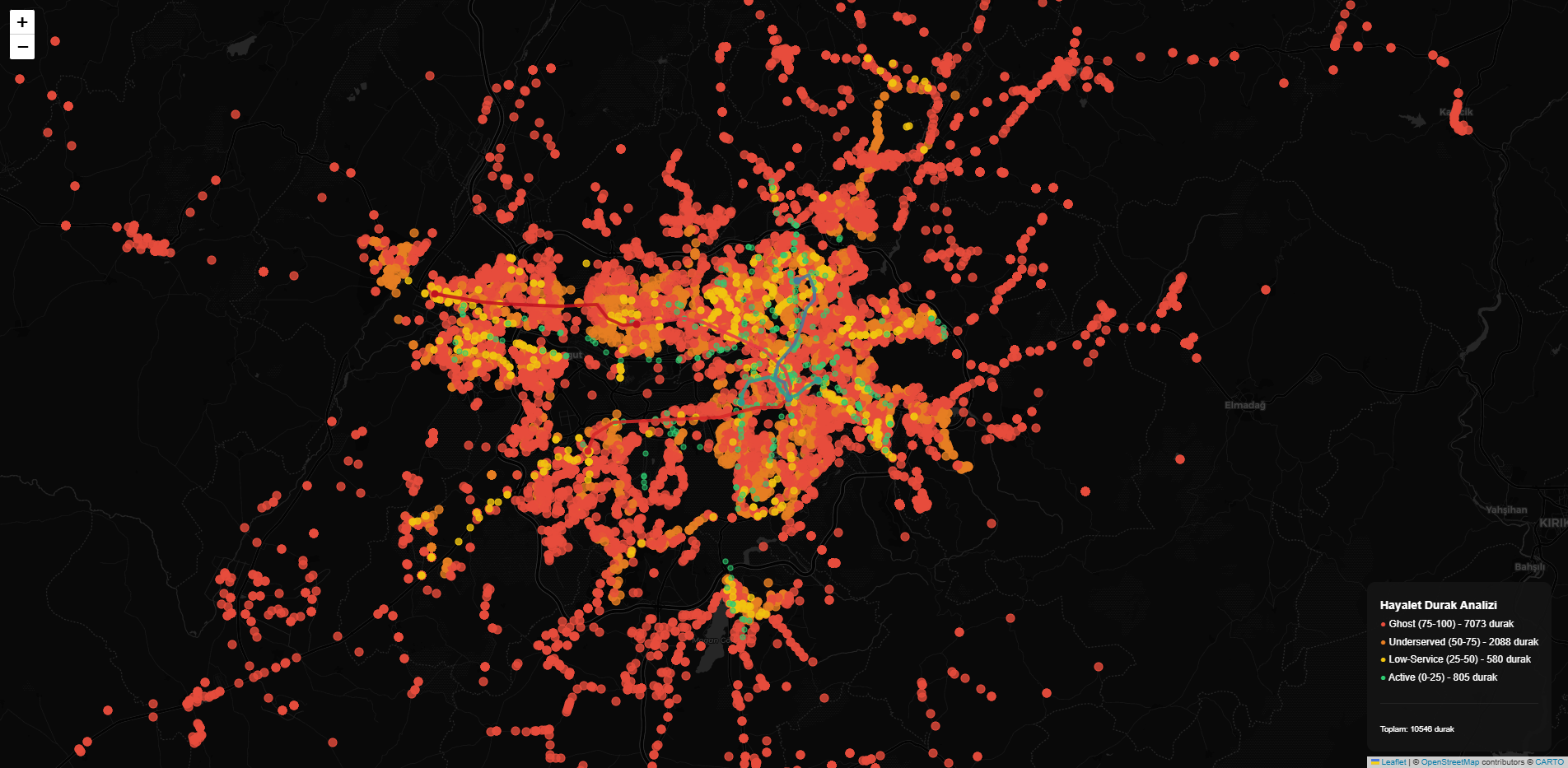}
    \caption{\textbf{Spatial distribution of "Ghost" stops}, identifying that 67.1\% of the network consists of physically present but operationally inactive nodes.}
    \label{fig:ghoststop}
\end{figure}

\hspace{1cm}This structural imbalance becomes even more evident through the \textit{"Ghost Stop"} Analysis in Figure \ref{fig:ghoststop}. Out of 10,546 total bus stops, 7,073 stops (67.1\%) are classified as \textit{"Ghost"} stops with scores between 75 and 100. An additional 2,088 stops (19.8\%) fall into the underserved category. Only 805 stops, representing 7.6 percent of the total, function as active service nodes. The average ghost score across the network is 77.1 out of 100, indicating that the majority of stops exist physically but operate with minimal trip frequency and passenger interaction. The threshold values are not arbitrary but based on the statistical distribution of Ghost Scores across the network. After calculating scores for all stops, the histogram of values showed natural clustering around four bands. These breaks were close to the quartile ranges of the dataset, so thresholds at 25, 50, and 75 were adopted to separate stops into low, moderate, high, and extreme service deficiency levels.

\begin{figure}
    \centering
    \includegraphics[width=1\linewidth]{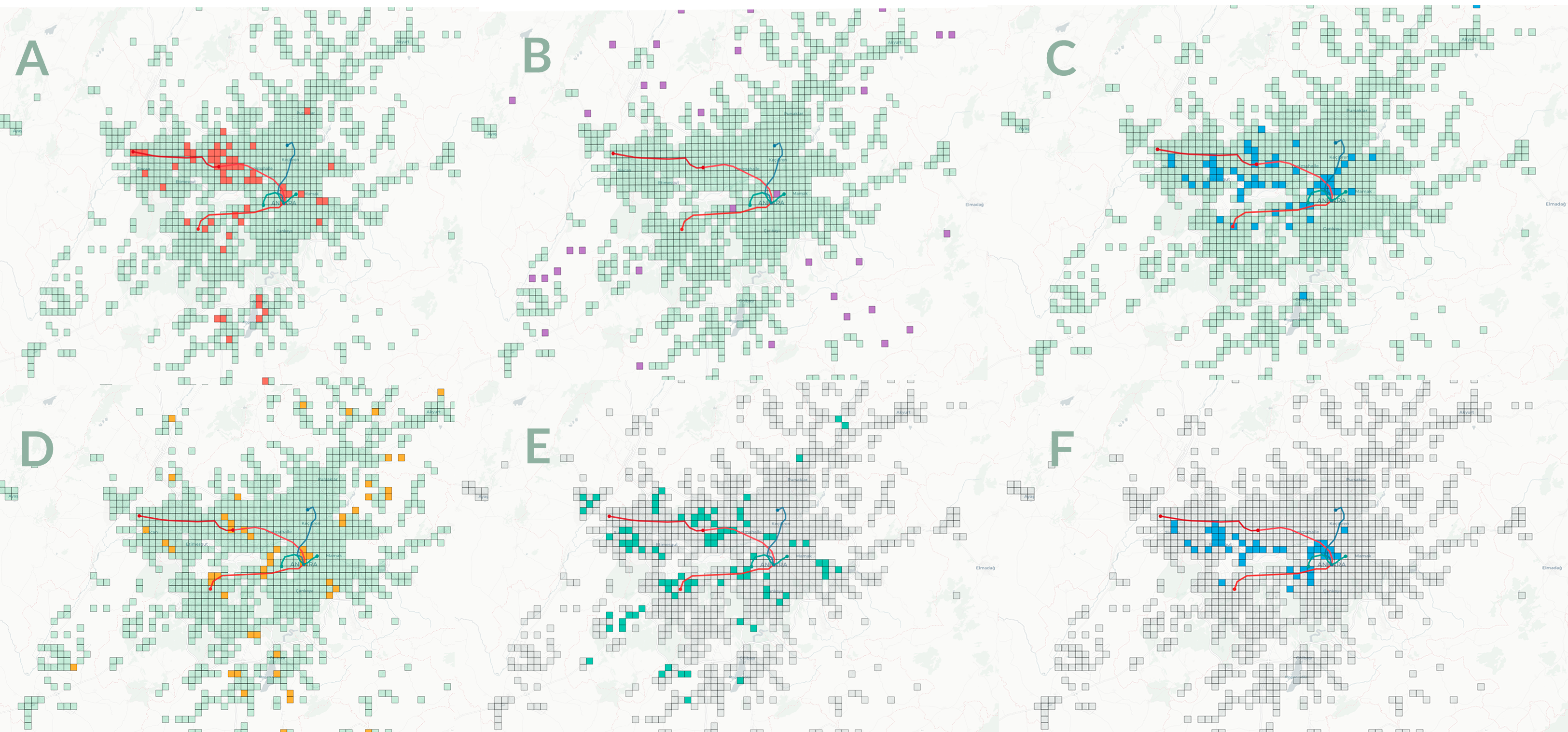}
    \caption{\textbf{Spatial distribution of anomalies and regime shifts identified by the six unsupervised algorithms:} (A) Autoencoder anomaly detection (red) based on reconstruction error; (B) Isolation Forest (blue) identifying statistical outliers; (C) Local Outlier Factor (orange) highlighting density-based contextual deviations; (D) Graph-based spatial anomaly (purple) showing divergence from geographic neighbors; (E) Embedding Stability Analysis (teal) visualizing high daily geometric drift in latent space; (F) Daily Clustering Regime Shift (green) indicating high frequency of categorical switching between operational archetypes.}
    \label{fig:6-grid}
\end{figure}

\hspace{1cm}The analytical framework is designed to diagnose the structural failures of Ankara's bus network through two complementary dimensions: \textit{spatial misalignment} and \textit{temporal instability} . Rather than relying on a single metric or algorithm, the framework employs an ensemble of \textit{six unsupervised methods, four for anomaly detection and two for regime shift analysis} in Figure \ref{fig:6-grid}, each capturing a mathematically distinct aspect of system failure. It is essential to distinguish between these two dimensions and their intersection. Service-demand mismatch, identified through the anomaly detection layer, is a static diagnosis: it flags grids where the time-averaged balance between ridership and service provision consistently deviates from the system norm—locations that are, in a sense, predictably underserved, and whose remedy is relatively straightforward. By contrast, a critical grid is a harder problem. It is a place where the gap between service and demand is large and also unstable over time. Demand and service change from day to day in unpredictable ways, so planners cannot clearly identify the problem or design a solution. Out of 1,382 analyzed grid cells, 46 exhibit service-demand mismatch, yet only 13 of these are simultaneously temporally unstable, qualifying as critical grids. These 13 locations show the most severe mismatch between centralized route planning and the spatiotemporal complexity of the city. In these zones, the transit network fails to meet demand and cannot clearly identify the pattern of that demand.

\hspace{1cm}The first analytical dimension deploys four complementary unsupervised anomaly detection algorithms, each designed to interrogate a distinct features of the service-demand relationship across Ankara's bus network in Table \ref{tab:anomaly-algorithms} and Figure \ref{fig:6-grid}. The first analytical dimension of the study is a Consensus Model composed of four complementary unsupervised anomaly detection algorithms designed to examine different aspects of the supply-demand relationship in Ankara’s bus network. The first algorithm, a PCA-based Autoencoder, treats the transit system as a low-dimensional manifold. It compresses a seven-dimensional feature space, including ridership, trips, and capacity, into three principal components, learns the latent statistical structure of the system, and attempts to reconstruct the data. Grids with high reconstruction error are flagged as structural mismatches that violate the network’s normal operational patterns. The second algorithm, Isolation Forest, detects global outliers by using random decision trees. A grid that can be isolated with a short path length is considered more anomalous, a property that helps identify leapfrog development areas in Ankara’s peripheral districts. The third method, Local Outlier Factor, shifts attention from global extremes to local context. It compares the density of each grid with the density of its twenty nearest neighbors and identifies contextual anomalies, locations that appear normal overall but are underserved relative to comparable areas. Finally, the graph-based spatial method models physical geography rather than abstract feature space. It constructs a neighborhood network within a 1.5 km radius and detects spatial breaks, grids that diverge sharply from their adjacent cells.

\begin{figure}
    \centering
    \includegraphics[width=1\linewidth]{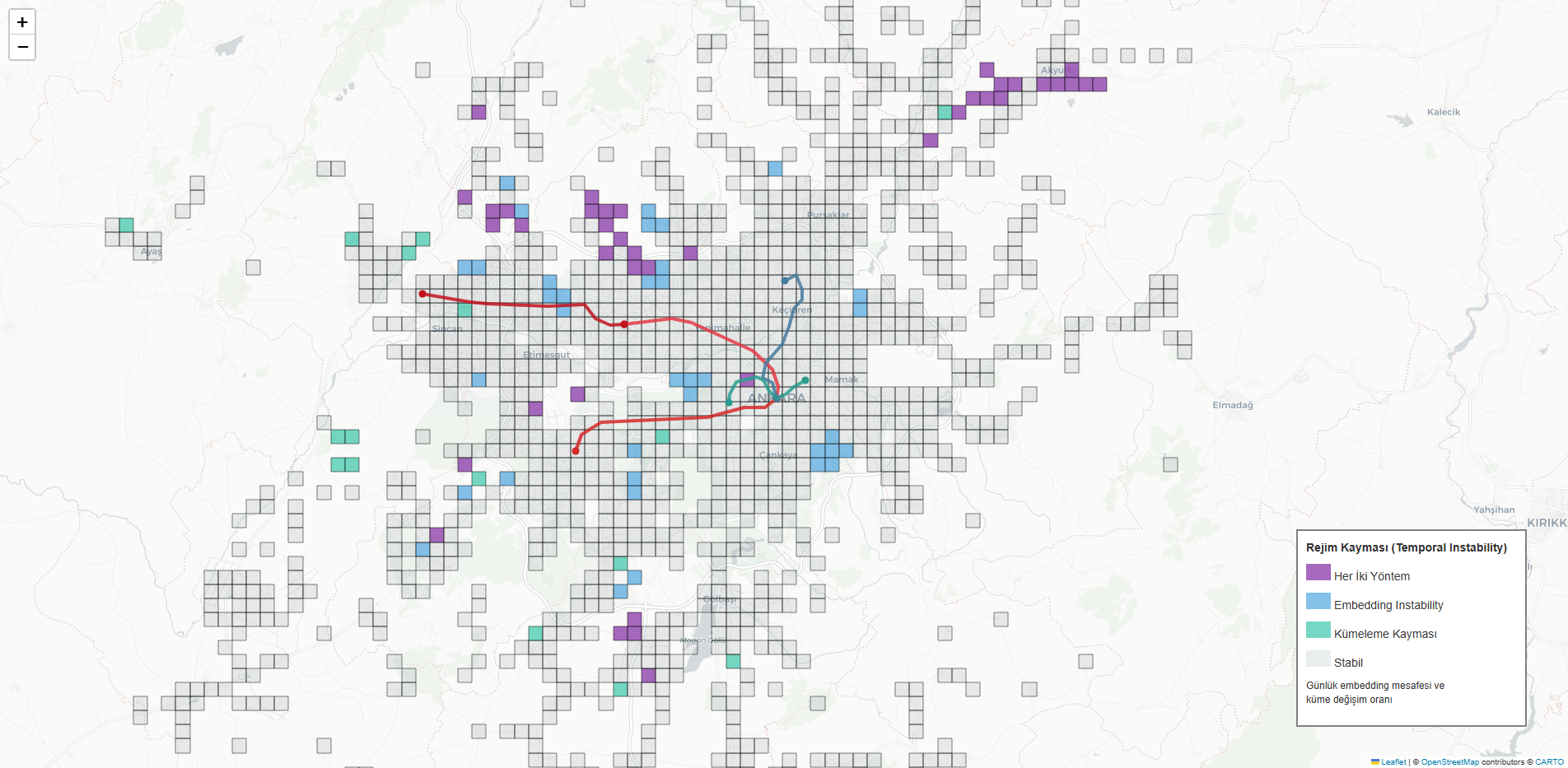}
    \caption{\textbf{Regime shift map} visualizing temporal instability: purple grids exhibit both geometric drift and frequent categorical switching, while blue and teal grids represent instability in only embedding space or cluster assignment, respectively.}
    \label{fig:regime-shift}
\end{figure}

\begin{figure}
    \centering
    \includegraphics[width=1\linewidth]{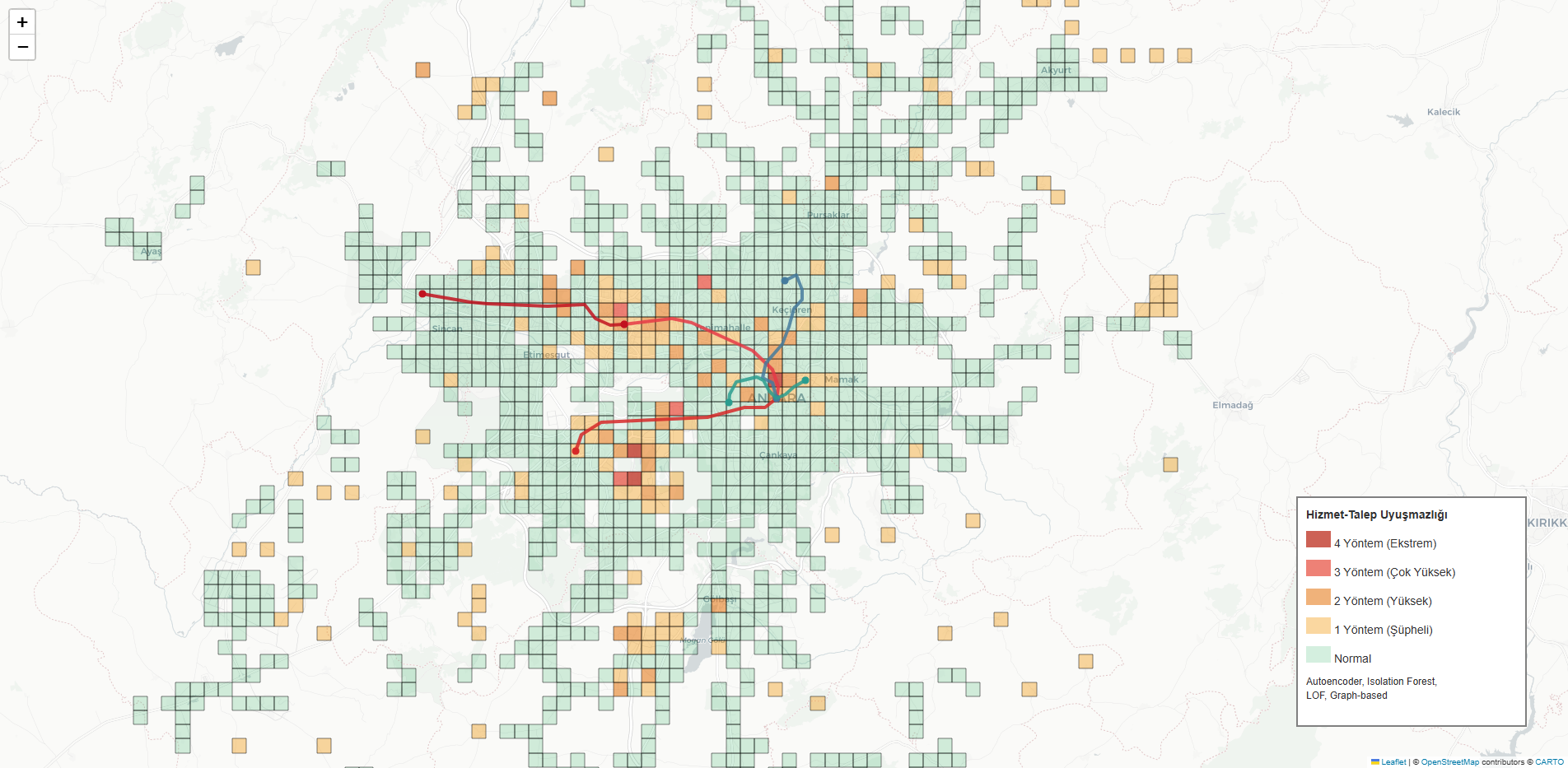}
    \caption{Anomaly map highlighting service-demand misalignment: red indicates critical anomalies confirmed by three or more detection methods, orange shows moderate anomalies confirmed by two methods, and green represents normal operating conditions.}
    \label{fig:anomaly-map}
\end{figure}

\begin{figure}
    \centering
    \includegraphics[width=1\linewidth]{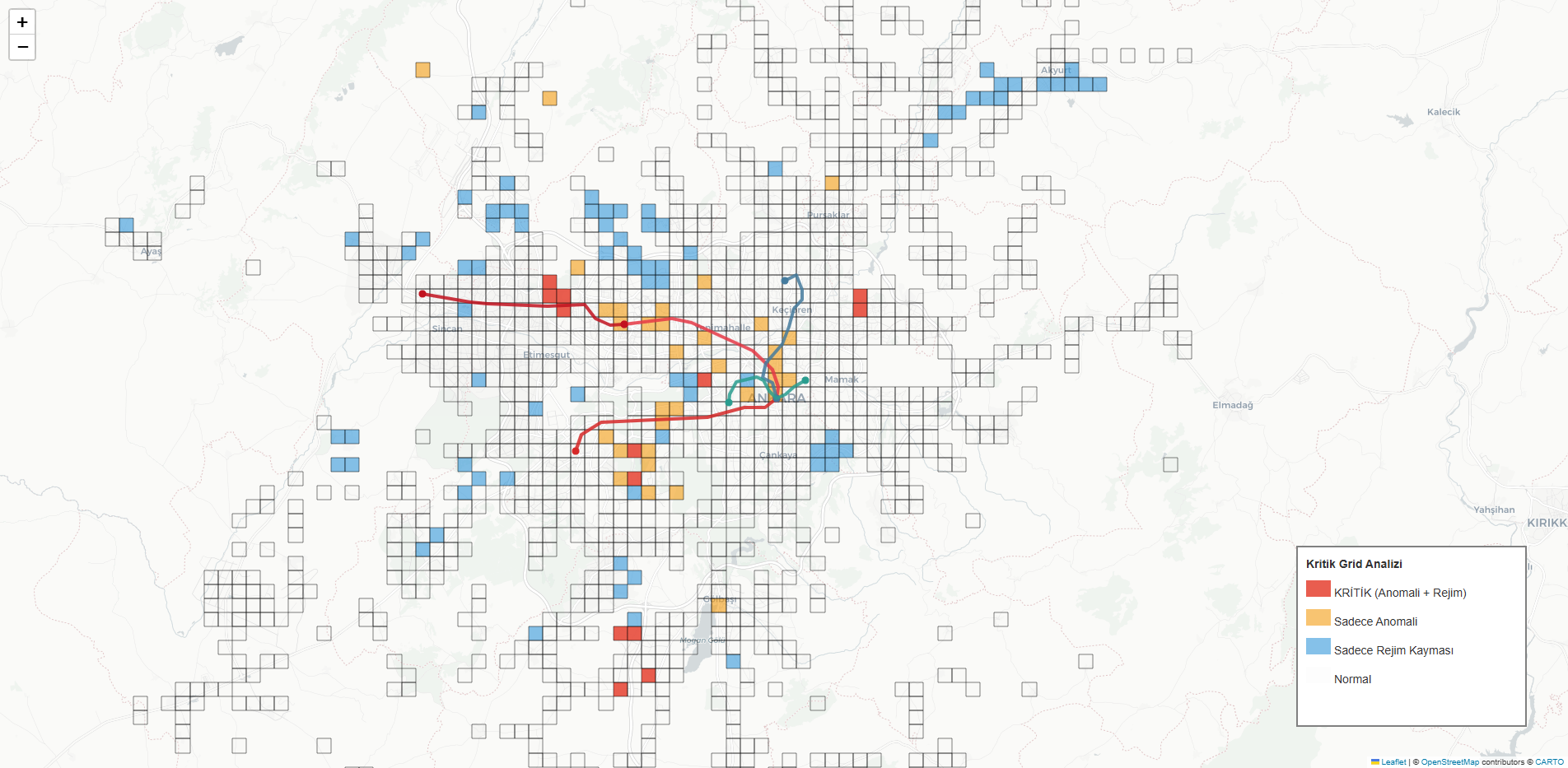}
    \caption{Spatial intersection of structural service-demand anomalies and temporal regime instability, highlighting 13 critical grid cells (red) where Ankara’s centralized transit planning fundamentally fails to accommodate the spatiotemporal complexity of urban sprawl.}
    \label{fig:criticalmap}
\end{figure}

\begin{table}[htbp]
\centering
\caption{Anomaly detection algorithms and their diagnostic perspectives on service-demand misalignment.}
\label{tab:anomaly-algorithms}
\small
\begin{tabular}{p{3cm} p{5.5cm} p{5.5cm}}
\toprule
\textbf{Algorithm} & \textbf{Diagnostic Question} & \textbf{Type of Failure Detected} \\
\midrule
Autoencoder (PCA) & Does this grid's ridership--frequency--occupancy relationship conform to the system's normative logic? & \textit{Systemic logic failure}: Locations whose supply-demand configuration cannot be reconstructed from the dominant latent patterns, indicating a breakdown in the system's internal grammar. \\
\addlinespace
Isolation Forest & Is this grid statistically extreme in the multidimensional feature space? & \textit{Extreme outliers}: Grids occupying the far edges of the distribution, such as exceptionally high passenger loads paired with minimal trip frequency. \\
\addlinespace
Local Outlier Factor & Does this grid deviate from its operational peer group? & \textit{Contextual neglect}: Locations that appear normal in absolute terms but are anomalously underserved relative to grids with comparable density profiles. \\
\addlinespace
Graph-based Spatial & Does this grid's service profile diverge from its geographic neighbors? & \textit{Spatial discontinuity}: Adjacent grid cells receiving dramatically different levels of service despite occupying the same urban fabric. \\
\bottomrule
\end{tabular}
\vspace{0.3em}
.
\end{table}

\begin{table}[htbp]
\centering
\caption{Regime shift algorithms and their diagnostic perspectives on temporal instability.}
\label{tab:regime-shift-algorithms}
\small
\begin{tabular}{p{3cm} p{5.5cm} p{5.5cm}}
\toprule
\textbf{Algorithm} & \textbf{Diagnostic Question} & \textbf{Type of Failure Detected} \\
\midrule
Embedding Stability & Does this grid's operational identity (ridership--frequency profile) remain consistent across days, or does it drift erratically? & \textit{Representational unpredictability}: Locations that behave like a busy urban center on one day and a quiet suburban periphery on the next, preventing the system from establishing a stable reading of the grid's identity. \\
\addlinespace
Daily Clustering & Can the system consistently assign this grid to the same operational archetype each day? & \textit{Categorical instability}: Locations whose assigned cluster oscillates between high-demand, low-service (deficit) and low-demand, low-service (normal) from one day to the next, indicating that the system cannot commit to a stable classification. \\
\bottomrule
\end{tabular}
\vspace{0.3em}

\end{table}

\hspace{1cm}The second analytical dimension expands the inquiry from the spatial detection of failure to the temporal assessment of its consistency in Figure \ref{fig:6-grid} and Table \ref{tab:regime-shift-algorithms}. It distinguishes between chronic deficiencies and erratic volatility using two metrics over the 173-day observation period. The first metric, \textit{Embedding Stability}, measures \textit{"identity drift"} by projecting each grid’s daily feature vector $(x_k(t))$ into a low-dimensional latent space using PCA and calculating the Euclidean distance $(d_k(t))$ between consecutive states. High drift values indicate locations that shift between very different operational identities, such as moving from a busy hub to a quiet periphery, which the central planning system cannot consistently interpret. The second metric, \textit{Daily Clustering Regime Shift}, measures categorical uncertainty. Each grid is assigned to an operational archetype using K-Means clustering, and the \textit{Clustering Switch Rate} $(SR_k)$ records how often a grid changes its classification.  

\hspace{1cm}Combining these temporal instability measures with spatial anomaly scores reveals three patterns of transit failure in Ankara in Figure \ref{fig:regime-shift}. \textit{Chronic Neglect} (high anomaly, low instability) describes areas where service is consistently poor but predictable and can be improved through capacity increases in low-frequency peripheral routes. \textit{Growing Pains of Sprawl} (low anomaly, high instability) appears in developing peripheral zones where demand fluctuates before stable commuter patterns form. \textit{Structural Collapse} (high anomaly, high instability) marks locations where the center-oriented bus network cannot meet demand or adapt to its variability, showing the strongest mismatch between Ankara’s expanding urban form and its transit structure.

\section{Conclusion}
Increasing the number of buses alone will not alter Ankara’s transport trajectory; it will only reinforce the existing \textit{“Skeletal Trap”}, a rigid, center-oriented transit structure unable to adapt to the city’s leapfrog urban expansion. The empirical evidence demonstrates that Ankara’s transport crisis is not fundamentally a shortage of vehicles but a crisis of architectural and structural mismatch. Capacity and trip distribution maps show that the network still follows a monocentric structure. Service is concentrated along the Kızılay–Ulus corridor and the main western routes, while scattered peripheral developments receive weak and irregular coverage. In this situation, adding more buses would mainly strengthen the existing radial pattern instead of solving the mismatch between urban growth and network design.

\hspace{1cm}A structural reconfiguration of the network is therefore necessary. While hub-and-spoke systems can function well in smaller or more compact cities, Ankara’s expanded urban form now change to a more mesh-based structure to solve dispersed travel patterns and reduce excessive dependence on central transfer corridors. Orbital and tangential bus corridors connecting metro lines outside the central core would eliminate enforced transfer dependency and reduce pressure on central hubs. Particular attention should be given to the underserved wedge zones located between metro corridors, where horizontal feeder lines can connect fragmented neighborhoods directly. Second, the high proportion of ghost stops demonstrates that traditional fixed-route operations cannot effectively serve leapfrog peripheral growth. \textit{Demand-Responsive Transit }systems should be implemented in low-density districts, allowing flexible routing that aligns service supply with real demand. Third, directional imbalance reflects a deeper morphological issue in which peripheral districts function primarily as dormitory suburbs. Encouraging mixed-use development and localized employment opportunities in areas such as İncek, Bağlıca, and western expansion corridors would reduce peak-hour commuter pressure and rebalance demand flows. Finally, terminal stations such as Koru and Törekent require synchronized multimodal integration. High-capacity feeder systems, including bus rapid transit or coordinated transfer scheduling, must be introduced to resolve the structural bottlenecks produced where metro capacity abruptly transitions to low-capacity buses.

\section{Limitations}
The study is constrained by several data limitations. The operational dataset provided only daily route-level aggregates, without hourly temporal resolution. As a result, peak-hour dynamics, directional passenger flows, and short-term congestion patterns could not be directly modeled. Furthermore, essential geospatial data such as stop coordinates, route geometries, and detailed network topology were not publicly available in open formats such as GTFS. After October 20, 2024, access to detailed operational data was also restricted, further limiting temporal continuity in the dataset. Because of these gaps, the analysis could not be conducted on an exact road-network basis. Instead, passenger distribution and service indicators were reconstructed using standardized 1 km × 1 km grid cells. This method reduces spatial ambiguity while avoiding artificial precision, yet it inevitably abstracts away micro-scale variations within individual routes and stops.

\hspace{1cm}Despite these constraints, the grid-based methodology provides a statistically consistent representation of system-level patterns. However, future research would benefit from access to high-resolution temporal data, detailed route geometries, and multimodal network information. Such datasets would enable dynamic simulation of passenger flows, more precise accessibility measurement, and more comprehensive evaluation of policy interventions within Ankara’s evolving urban morphology.

\bibliographystyle{unsrt}  
\bibliography{references}

\end{document}